\documentclass[twocolumn]{aastex631}
\graphicspath{{./}{figure/}}
\usepackage{natbib}
\usepackage{lineno}
\usepackage{comment}

\begin{document}

%\bibliographystyle{aasjournal}
%\begin{thebibliography}

\title{Environments around quasars at $z\sim$3 revealed by wide-field imaging with Subaru HSC and CFHT}

\author[0009-0000-2525-9236]{Yuta Suzuki}
\affiliation{Graduate School of Science and Engineering, Ehime University, 2-5 Bunkyo-cho, Matsuyama, Ehime 790-8577, Japan; yuta@cosmos.phys.sci.ehime-u.ac.jp}

\author[0000-0002-0673-0632]{Hisakazu Uchiyama}
\affiliation{National Astronomical Observatory of Japan, 2-21-1 Osawa, Mitaka, Tokyo 181-8588, Japan}\affiliation{Research Center for Space and Cosmic Evolution, Ehime University, 2-5 Bunkyo-cho, Matsuyama, Ehime 790-8577, Japan}

\author[0000-0001-5063-0340]{Yoshiki Matsuoka}
\affiliation{Research Center for Space and Cosmic Evolution, Ehime University, 2-5 Bunkyo-cho, Matsuyama, Ehime 790-8577, Japan}

\author[0000-0001-5394-242X]{Jun Toshikawa}
\affiliation{Nishi-Harima Astronomical Observatory, Center for Astronomy, University of Hyogo, Sayo, Hyogo 679-5313, Japan}
\affiliation{Department of Physics, University of Bath, Claverton Down, Bath, BA2 7AY, UK}

\author[0000-0001-8221-8406]{Stephen Gwyn}
\affiliation{NRC-Herzberg, 5071 West Saanich Road, Victoria, British Columbia V9E 2E7, Canada}

\author[0000-0001-6186-8792]{Masatoshi Imanishi}
\affiliation{National Astronomical Observatory of Japan, 2-21-1 Osawa, Mitaka, Tokyo 181-8588, Japan}
\affiliation{Department of Astronomy, School of Science, Graduate University for Advanced Studies (SOKENDAI), Mitaka, Tokyo 181-8588, Japan}
\affiliation{Toyo University, 5-28-20, Hakusan, Bunkyo-ku, Tokyo 112-8606, Japan}

\author[0000-0002-4718-3428]{Chengze Liu}
\affiliation{Department of Astronomy, School of Physics and Astronomy, and Shanghai Key Laboratory for Particle Physics and Cosmology, Shanghai Jiao Tong University, Shanghai 200240, People's Republic of China}

\author[0000-0002-5197-8944]{Akatoki Noboriguchi}
\affiliation{Center for General Education, Shinshu University, 3-1-1 Asahi, Matsumoto, Nagano 390-8621, Japan}

\author[0000-0002-7712-7857]{Marcin Sawicki}
\affiliation{Institute for Computational Astrophysics and Department of Astronomy and Physics, Saint Mary\'s University, 923 Robie Street, Halifax, Nova Scotia, B3H 3C3, Canada}

\author[0000-0002-3531-7863]{Yoshiki Toba}
%\altaffiliation{NAOJ fellow}
\affiliation{National Astronomical Observatory of Japan, 2-21-1 Osawa, Mitaka, Tokyo 181-8588, Japan}
\affiliation{Research Center for Space and Cosmic Evolution, Ehime University, 2-5 Bunkyo-cho, Matsuyama, Ehime 790-8577, Japan}
\affiliation{Academia Sinica Institute of Astronomy and Astrophysics, 11F of Astronomy-Mathematics Building, AS/NTU, No.1, Section 4, Roosevelt Road, Taipei 10617, Taiwan}

%\author{et al. }

\begin{abstract}
We examine the local density environments around 67 quasars at $z\sim3$, by combining the imaging data of Hyper Suprime-Cam Subaru Strategic Program (HSC-SSP) and Canada-France-Hawaii Telescope Large Area U-band Survey (CLAUDS) over about 20 deg$^{2}$. %and Sloan Digital Sky Survey (SDSS).
%We measure the local densities of
Our measurements exploit $U$-dropout galaxies in the vicinities of quasars taken from Sloan Digital Sky Survey (SDSS).
We find that the quasars have the %same
indistinguishable surrounding density distribution from the $U$-dropout galaxies, and that three quasars are associated with protocluster candidates within a projected separation of 3 arcmin.
%overdense regions within three arcmin.
According to a halo evolutionary model, our results suggest that quasars at this epoch occupy haloes with a typical mass of $1.3^{+1.4}_{-0.9} \times 10^{13} h^{-1} \mathrm{M_{\odot}}$.
We also investigate the dependence of the local galaxy overdensity on ultraviolet (UV) luminosities, black hole (BH) masses, and proximity zone sizes of the quasars, but no statistically-significant correlation was found.
%As a result, all of them are independent of the surrounding densities.
%The quasars host the most massive BHs tend to avoid the most overdense regions.
%No significant correlation is found the the proximity zone sizes and the local densities.
%To resolve the proximity effcts, 
%We also make a radial density profile around the quasars.
Finally, we find that the local density of faint $U$-dropout galaxies are lower than that of bright $U$-dropout galaxies within a projected distance of $0.51\pm0.05$ physical Mpc, where the quasar UV radiation is 30 times intenser than backgroung UV radiation. 
%According to the photoevaporation model,
We argue that photoevaporation may suppress galaxy formation at short distances where the quasar UV intensity is strong, even in massive haloes.
\end{abstract}

%\linenumbers
%%% Introduction
\section{Introduction} \label{sec:intro}
Quasars are one of the most luminous objects in the Universe. %by gas accretion onto supermassive black holes (SMBHs) at the centers of host galaxies.
The strong ultraviolet (UV) radiation from quasars is expected to ionize gas and suppress star formation in not only the host but also the surrounding haloes \citep[``photoevaporation'';][]{Benson2002, Kashikawa2007, Utsumi2010, Simpson2014, Kikuta2017, Uchiyama2019}.
In order to fully understand galaxy formation and evolution, we thus need to characterize where quasars appear in the large scale structure of galaxies, both observationally and theoretically. %observationally.  %On the other hand, 

Clustering analysis is a good approach to measure the average halo environments of quasars.
At $z<3$, quasars are found to reside in the same environments as do galaxies, with the quasar minimum dark matter (DM) halo masses estimated to be $\thicksim10^{11.5-12}h^{-1}\mathrm{M_{\odot}}$ \citep{Myers2006, Ross2009, White2012, Font-Ribera2013}.
These DM halo masses are not the most massive haloes nor the progenitors of the most massive local clusters at $z<3$ \citep[e.g.,][]{Fanidakis2013}. %The local galaxy densities of quasars are also reported to be similar to that of galaxies \citep[e.g.][]{Karhunen2014, Alam2021}.
 
At $z\sim3$ and beyond, quasar environments are still unclear.
\citet{Shen2007} found a significantly larger minimum DM halo mass of $(2-3)\times10^{12}h^{-1}\mathrm{M_{\odot}}$ for quasars taken from % which is significantly larger than that of \citet{Eftekharzadeh2015} using a luminous quasar sample ($M_{i}\lesssim26.5$)
Sloan Digital Sky Survey \citep[SDSS:][]{York2000}. 
It indicates that a significant fraction of the quasar haloes at $z\sim3$ evolve into present-day massive galaxy clusters.
On the other hand, \citet{Eftekharzadeh2015} used the data from Baryon Oscillation Spectroscopic Survey \citep[BOSS:][]{Dawson2013} and found a minimum DM halo mass of $(0.60-0.72)\times10^{12}h^{-1}\mathrm{M_{\odot}}$ for quasars at $z\sim3$, which is comparable to those found at $z<3$.
 %In the case of the DM halo mass estimated by \citet{Shen2007}, it is comparable to that of a present-day massive galaxy cluster.
 %While that of \citet{Eftekharzadeh2015} is lower with an order of magnitude than the DM halo mass that would be expected if protocluster were traced \citep[see][for a comprehensive review of protocluster]{Overzier2016}.
The discrepancy between the halo mass measurements by \citet{Shen2007} and \citet{Eftekharzadeh2015} remains controversial.

Another method to quantity the environments is to measure galaxy number density around quasars.
\citet{Falder2011} investigated the environments of 46 luminous quasars ($M_{i}\lesssim-26$) at $0.3<z<5.3$ using Spitzer Space Telescope/Infrared Array Camera \citep[IRAC:][]{Fazio2004}.
11 quasars at $z\sim3$ were found to live in $\sim2\sigma$ overdense regions on average. 
%This limited sample may cause a large scatter of the surrounding densities.
%In order to make the result more significant, it is necessary to construct a larger sample. %(at least four times).
% by stacking 11 sub-samples.
\citet{Fossati2021} examined 27 quasar environments by using Very Large Telescope (VLT)/Multi Unit Spectroscopic Explorer \citep[MUSE:][]{Bacon2010}.
They found that the number density of Ly$\alpha$ Emitters (LAEs) around quasars is higher at $3.6 \, \sigma$ significance than in blank fields.
However, the galaxy sample of \citet{Falder2011} has a relatively shallow  $5\sigma$ limiting magnitude of 24.5 in the $r$-band, which has missed numerous faint objects.
\citet{Fossati2021} covers a relatively small area of about 0.30 Mpc around the quasar, while a larger area needs to be investigated to study the relationship between large-scale structures and quasars.
% the previous studies are not yet sufficient in providing a statistically-significant number of quasars or in probing a large enough volume of the underlying large-scale structure.
%In this study, however, the spatial coverage around quasars is limited ($\sim10^{2} \mathrm{kpc}$). 
%The photoevaporation effect is expected to extend to several Mpc \citep[e.g.][]{Kashikawa2007}. Thus it is necessary to investigate a wide range of area where clustering and photoevaporation effects can be disentangled.
%requiring an larger area to be investigated because the effect of quasar clustering and photoevaporation may not be separable.
An additional complexity comes from the fact that galaxies, in particular low-mass galaxies including LAEs can be affected by photoionization effects from the strong quasar UV radiation \citep[e.g.][]{Kashikawa2007, Cantalupo2012}.
It reduces the number of galaxies in the vicinities of quasars, and thus, conceals a signal on where quasars are preferentially formed. 
In order to overcome these problems, it is necessary to exploit deep and wide observations of galaxies and study the environment in which quasars exist.

In this study, we present a statistical measurement of the local density environments of quasars at $z\sim3$, by using the imaging data from Hyper Suprime-Cam Subaru Strategic Program \citep[HSC-SSP;][]{Aihara2018a, Aihara2018b, Aihara2019, Aihara2022} and Canada-France-Hawaii Telescope (CFHT) Large Area U-band Deep Survey \citep[CLAUDS;][]{Sawicki2019}. %is newly released \citep[][]{}. 
%can provide us with an unprecedently large sample of Lyman Break Galaxies \citep[LBGs;][]{Ono18} which are less effective to the photoionization effects \citep{Kashikawa2007}. 
The combined data allow us to construct a large sample of Lyman Break Galaxies (LBGs) at $z\sim3$, which are less affected by the photoionization effects than are LAEs \citep{Kashikawa2007}. 
The overdensity map of the LBGs has been constructed from the HSC-SSP and CLAUDS data \citep{Toshikawa2024}, in the same manner as in \citet{Toshikawa2016, Toshikawa2018}.
We measure the overdensity significances around quasars to assess the halo mass at $z\sim3$ \citep[][]{Shen2007, Eftekharzadeh2015}, and interpret the results using the Extended Press Schechter model \citep[][]{Bond1991, Bower1991, Lacey1993}. 
Possible correlations between quasar properties and their surrounding densities are also examined, which may signal 
%This analysis can shed light on
the existence of the photoevaporation effect on the surrounding galaxies \citep[][]{Uchiyama2018}. 

%the lying baryonic physics quasar environments at $z\sim3$ through this analysis.
%present the results of the association between densities of LBGs and quasars at $z\sim3$, using the data of HSC-SSP, CLAUDS, and SDSS.

The structure of this paper is as follows.
Section 2 describes the HSC-SSP and CLAUDS data, construction of the overdensity map of LBGs, %constructed by Toshikawa et al. (in prep) using the HSC-SSP + CLAUDS data,
and the quasar sample taken from SDSS.
We present a method to quantify the relationship between quasars and their local densities in Section 3.
In Section 4, we present the results of the measured quasar environments and their possible correlation with quasar properties. 
% correlation between the density of galaxies and quasars.
In Section 5, we discuss the possible implications of the results.
Finally, we present our conclusions in Section 6. 

In this paper, magnitudes are given in the AB system \citep{Oke1983, Fukugita1995}.
We adopt a Lambda cold dark matter ($\Lambda$CDM) cosmology with $H_{0}=70~\mathrm{km\,s^{-1}\,Mpc^{-1}}$, $\Omega_{M}=0.3$, and $\Omega_{\Lambda}=0.7$, which are consistent with the values reported by \citet{PlanckColabo2020}.
This study also uses cModel magnitude, which is measured by fitting two-component, PSF-convolved galaxy morphology model to a given object profile \citep{Abazajian2009}.%, for photometry.

%%% Data and sample selection
\section{Data and sample selection} \label{sec:data}

\subsection{HSC-SSP and CLAUDS} \label{subsec:hsc}
%HSC-SSP and CLAUDS are embraced in this study \citep[for details, see][respectively]{Aihara2018, Sawicki2019}. % These surveys are described in \citep[HSC-SSP]{Aihara2018} and \citep[CLAUDS]{Sawicki2019} in detail.
%Here is a brief explanation. 
The HSC-SSP survey \citep{Aihara2018a} started in 2014 and finished in 2021.
It used a wide-field camera, HSC, installed on Subaru Telescope.
%HSC-SSP provides imaging data in $g, r, i, z$, and $y$ band filters using Subaru Telescope/HSC \citep{Kawanomoto2018}.
HSC consists of 116 2K$\times$4K CCDs built by Hamamatsu Photonics K.K., Japan.
Of these CCDs, 104 are used to obtain scientific data in a field of view of 1.5 degrees in diameter.
The detailed system designs are described in \citet{Miyazaki2018} and \citet{Komiyama2018}.
The HSC-SSP survey consists of three layers, i.e., Wide, Deep, and UltraDeep, with different combinations of the area and depth of the observations.
%The DR16A data we use in this study \citep{Aihara2018b} 
HSC-SSP provides imaging data in $g, r, i, z$, and $y$ bands as well as a few narrow-band filters \citep{Kawanomoto2018}.
The data were reduced by hscPipe \citep{Bosch2018}, a software developed based on the Large Synoptic Survey Telescope pipeline \citep{Ivezic2008, Axelrod2010}.
The astrometric and photometric calibration are tied to the Panoramic Survey Telescope and Rapid Response System (Pan-STARRS) 1 system \citep{Schlafly2012, Tonry2012, Magnier2013}. 
The present work is based on the HSC-SSP Deep/UltraDeep layer data contained in the S16A internal data release which corresponds to public data release 1 \citep{Aihara2018b}.
 % The galaxy sample is obtained from S16A internal data \citep{Aihara2018b}.% 今回のデータはS16Aであることを書く+reference

The CLAUDS \citep{Sawicki2019} observation provides the $U$-band imaging data with the CFHT/MegaCam. % \citep{Sawicki2019}.
 %In CLAUDS used both the old $u^{\ast}$ and new $u$ filter,because
The $U$-band images are obtained with two filters, the old $u^{\ast}$ and new $u$ filter, because the MegaCam filter set was replaced during the survey.
MegaCam consists of 40 2K$\times$4K CCDs built by CEA Saclay, France.
The CCDs cover a field of view of 1$\times$1 square degree.
The system designs are described in \citet{Boulade2003} in detail.
%It is a very complementary survey; the HSC-SSP Deep survey and UltraDeep survey fields were completely covered by the $u$-band survey.

CLAUDS covers a large part of HSC-SSP Deep and UltraDeep layers.
The $u$-band ($u^{\ast}$-band) data are available in E-COSMOS, COSMOS, DEEP2-3, and ELAIS-N1 (XMM-LSS and SXDS). 
The four Deep fields (E-COSMOS, XMM-LSS, ELAIS-N1, DEEP2-3) and the two UltraDeep fields (COSMOS, SXDS) cover the total area of 17.32 $\mathrm{deg}^{2}$ with the $5\sigma$ limiting magnitudes of $(U,g,r,i,z,y)_{\mathrm{Deep}}=(27.1, 26.8, 26.6, 26.5, 25.6, 24.8)$ and $(U,g,r,i,z,y)_{\mathrm{UltraDeep}}=(27.7, 27.4, 27.3, 27.0, 26.4, 25.6)$ measured in 2 arcsec apertures \citep{Aihara2018b, Sawicki2019}. 
 % 何の種類の等級か S16Aでの値にする
%The two UltraDeep survey fields  cover the area of 1.54 $\mathrm{deg}^{2}$ with .
% and the  data obtained in .
\citet{Sawicki2019} describes the joint processing of the CLAUDS and HSC-SSP data, which we use in the present analysis.
 
\subsection{Overdensity map of $U$-dropout galaxies} \label{subsec:overdensity}
We use the overdensity map of $U$-dropout galaxies created by \citet{Toshikawa2024}.
The method of constructing the map is described in detail in \citet{Toshikawa2016}.
We give a brief summary below.

First, $U$-dropout galaxies are selected from the six Deep and UltraDeep fields using the following color criteria \citep{Sawicki2019}:

For $u$-dropouts
\begin{equation}
g-r<1.2,
\end{equation}
\begin{equation}
u-g>0.88,
\end{equation}
\begin{equation}
u-g>1.88(g-r)+0.68,
\end{equation}

and for $u^{\ast}$-dropouts
\begin{equation}
g-r<1.2,
\end{equation}
\begin{equation}
u^{\ast}-g>0.9,
\end{equation}
\begin{equation}
u^{\ast}-g>1.5(g-r)+0.75.
\end{equation}
The typical redshift distribution of galaxies selected with these criteria is $z\sim2.6-3.6$ \citep{Toshikawa2024}.

Next, the number of $U$-dropout galaxies is counted within an aperture distributed over the fields. % at spatial interval of 20 arcsec. 
The aperture size is 1.6 arcmin corresponding to 0.75 proper Mpc (pMpc) which is the typical protocluster radius at $z\thicksim3.1$ \citep[e.g.,][]{Chiang2013, Toshikawa2016}.
The overdensity significance is defined as
\begin{equation}
\sigma = \frac{N-\bar{N}}{\sigma_{N}},
\end{equation}
where $N$ is the number of $U$-dropout galaxies in the aperture. $\bar{N}$ and $\sigma_{N}$ represent the mean and standard deviation of the galaxy numbers measured in each of the six fields.
 %in the aperture, $\bar{N}$ is the average of $N$, and $\sigma_{N}$ is the standard deviation of $N$, in each field.

From the overdensity map, protocluster candidates were selected as $>4\sigma$ overdense regions.
\citet{Toshikawa2016} showed that $\sim76 \pm 15\%$ of such candidates are expected to be real protoclusters with halo mass evolving to $>10^{14} \mathrm{M_{\odot}}$ at $z=0$.
As a result, 24 protocluster candidates were found in the present work.
The completeness of finding protocluster from our overdensity map is estimated to be $\sim5 \pm 1\%$ \citep[][]{Toshikawa2016}.

%%% 
\subsection{Quasar sample} \label{subsec:quasar}
Our quasar sample is extracted from the SDSS DR16 QSO catalog (DR16Q) \citep{Lyke2020}. 
The catalog contains 750,414 quasars. % used in Toshikawa et al. (in prep).
We select the quasars whose redshift range corresponds to the redshift distribution of the $U$-dropout galaxies, where the selection completeness is higher than 50\% \citep{Toshikawa2016, Toshikawa2024}.
We found 67 quasars at $z=2.8-3.4$ in the Deep and UltraDeep survey fields, and visually inspected all the spectra to confirm that they are indeed quasars.
% As the redshift distribution of the quasars is slightly narrower than that of the $U$-dropout galaxies, we aim to determine not whether the $U$-dropout overdense regions host the quasars, but whether the number density environments around the quasars are overdense regions.
Among the 67 quasars, 45 quasars are above the completeness flux limit of the eBOSS selection, and constitute the ``complete sample'' which will appear later.
The complete limit corresponds to the eBOSS limiting magnitude of $g < 22$ mag or $r < 22$ mag \citep{Myers2015}. %, and assuming that $r - i = 0.0954$ which is the median color of the our quasar sample.
Assuming a quasar continuum slope of $f_{\mathrm{\nu}} \propto \nu^{-0.5}$ \citep{VandenBerk2001} and the median C {\footnotesize IV} width of our sample $\mathrm{FWHM} = 4300 \, \mathrm{km \, s^{-1}}$ (see Section 3.2.),
the above completeness limit corresponds to $M_{i} = -25.5$ and $\log ({M_{\mathrm{BH}}/\mathrm{M_{\odot}}) = 8.4}$.

\section{Methods} \label{sec:methods}
\subsection{Measurement of environment}
We measure the overdensity significances in three scales.
The overdensity at the exact position where a quasar is located is called the ``Nearest'' overdensity, hereafter.
On the other hand, 
%at which the quasars are located (we call ``Nearest'' hereafter). 
%The
quasars are not always found at the center of overdense regions \citep{Venemans2007}. 
We thus measure the maximum overdensity significances within 1 and 3 arcmin from a quasar (hereafter ``\textless~1 arcmin'' and ``\textless~3 arcmin'', respectively).
Note that \citet{Chiang2013} found that the typical radius of protoclusters, whose host haloes evolve to $>10^{14}\mathrm{M_{\odot}}$ at $z=0$, is $\lesssim 3.0$ arcmin at $z=3$.
At this redshift, 1 and 3 arcmin corresponds to 0.47 and 1.4 pMpc, respectively.
% u-dropoutの銀河周辺の密度もクエーサーと同様の手法で測る
For comparison, the overdensity significances around the $U$-dropout galaxies themselves are also measured with the same method as for the quasars.

\subsection{Measurement of quasar properties} \label{subsec:measurequasar}
The $i$-band absolute magnitudes at $z=2$, {\tt\string M\_I}, is taken from the SDSS DR16 catalog. 
{\tt\string M\_I} corresponds to the rest-frame UV luminosity for objects at $z\sim3$ \citep[for detail, see][]{Richards2006, Lyke2020}. 
%It should be noted that they were corrected for extinctions and $K$ corrections were normalized to a redshift 2 \citep[for detail,][]{Lyke2020}.
 %The absolute magnitude corresponds to the rest wavelength of $2500 \mathrm{\AA}$ \citep{Richards2006}.

%\subsubsection{BH mass} \label{subsec:tables}
%To estimate the black hole (BH) mass, we use the reduced one-dimensional spectral data of the quasars, which are available in the SDSS Science Archive Server (SAS).
Of the 67 quasars, 62 has BH mass measured and reported in SDSS DR14 catalog \citep{Rakshit2020}.
%The BH masses of the remaining five quasars are measured by
%taken from SDSS DR14 catalog \citep[][]{Rakshit2020}.  
%The BH masses were estimated by \citet{Rakshit2020}. 
%We find that in addition to the SDSS DR14 quasars, five SDSS DR16 quasars reside in our survey region. 
We estimate the BH masses of the remaining five quasars by using the single-epoch virial BH mass estimator \citep[e.g.,][]{Shen2013};
%For the five newly added SDSS DR16 quasars within the six HSC-SSP regions, we estimate BH masses. % 加えたBH質量について書く
%The value of quasar BH mass can be estimated by the single-epoch virial BH mass estimator \citep{Shen2013}; %This estimator takes the form of

\begin{eqnarray}
 \log \biggl(\frac{M_{\mathrm{BH}}}{\mathrm{M_{\odot}}}\biggr) = \mathrm{A} + \mathrm{B}\log \biggl(\frac{L}{10^{44}~\mathrm{erg~s^{-1}}}\biggr) \nonumber\\
 + \mathrm{C}\log \biggl(\frac{\Delta V}{\mathrm{km~s^{-1}}}\biggr),
\end{eqnarray}
where $L$ [erg/s] is the quasar continuum luminosity at $1350 \, \mathrm{\AA}$ and $\Delta V$ $\mathrm{[km~s^{-1}]}$ is the full width at half maximum (FWHM) of an emission line.
For $z\sim3$ quasars, C \footnotesize{I\hspace{-1pt}V} \normalsize line is the common emission line to estimate the BH mass on single-epoch measurement, because the other lines do not fall into the SDSS spectral coverage.
When using the above equation with  C \footnotesize{I\hspace{-1pt}V} \normalsize line, the coefficients A, B, and C are 0.660, 0.53, and 2, respectively \citep{Vestergaard2006}.
The same procedure was also used in \citet{Rakshit2020}.
We subtract the best-fit continuum component from the quasar spectra with a single power law determined in the wavelenths range of $1445-1455$\,\AA\, and $1695-1705$\,\AA, where no strong emission lines are present \citep{Vestergaard2006}.
%We subtract the best-fit continuum component from an observed spectrum,
Then % from the quasar spectrum. %continuum of the spectrum at 1445-1455~\AA~and 1695-1705\AA~without mixing emission lines.
the C \footnotesize{I\hspace{-1pt}V} \normalsize line width $\Delta V$ is estimated by fitting a Gaussian model.

\begin{figure}[h!]
\begin{center}
\includegraphics[width=1.0\linewidth]{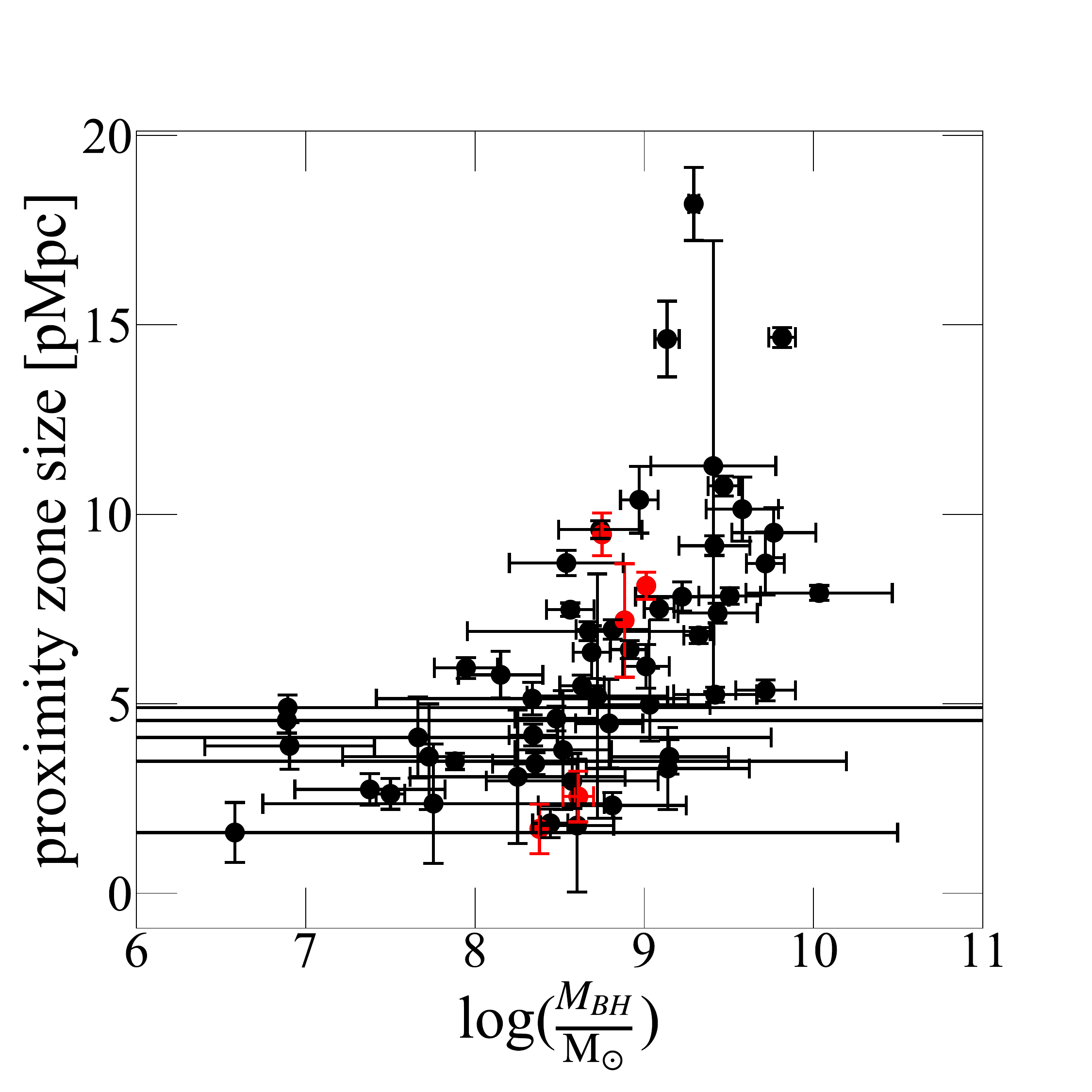}
%\makeblank{1}
\end{center}
\vspace*{-0.7cm}
\caption{The relation between proximity zone sizes and the black hole masses of the 67 quasars.
The black dots indicate quasars taken from the DR14 catalog of \citet{Rakshit2020}, and the red dots indicate the DR16 quasar for which we measured BH masses.} \label{prop}
\end{figure}

%\subsubsection{Proximity zone size of the quasar} \label{subsec:tables}
 We also measure proximity zone size with essentially the same method as described in \citet{Calverley2011}.
The photoionization rate of H \small I \normalsize by quasar radiation is given by
\begin{equation}
\Gamma_{\mathrm{Q}}=\int_{\nu_{\mathrm{L}}}^\infty \frac{4\pi J_{\mathrm{\nu}}\sigma_{\mathrm{H_{I}}}(\nu)}{\mathrm{h}\nu}d\nu \label{gamma},
\end{equation}
where $\nu_{\mathrm{L}}$ is the frequency at the Lyman limit, $J_{\mathrm{\nu}}$ is the intensity of quasar radiation, $\sigma_{\mathrm{H_{I}}} (\nu) = 6.3\times10^{-18}$ $(\nu_{\mathrm{L}}/\nu)^{2.75}$ $\mathrm{cm}^{2}$ is the ionization cross-section of H \small I \normalsize, and $\mathrm{h}$ is the Planck constant.
Assuming a single power-law $(f_{\mathrm{\nu}} \propto \nu^{-\alpha})$ for quasar radiation, equation (\ref{gamma}) becomes
\begin{equation}
\Gamma_{\mathrm{Q}} = \frac{9.5 \times 10^{8}F_{\nu_{\mathrm{L}}}}{\alpha+2.75} \label{gammaf},
\end{equation}
where $F_{\mathrm{\nu_{L}}}$ is the flux density at the Lyman limit.
%The continuum is fitted at the wavelength windows of $1700-1900$ \AA ~and $960-1100$ \AA.
On the other hand, $F_{\mathrm{\nu_{L}}}$ is expressed as % $F_{\mathrm{\nu_{L}}}(R_{\mathrm{pz}})$ using
\begin{equation}
F_{\mathrm{\nu_{L}}}(R) = \frac{L_{\mathrm{\nu_{L}}}}{4\pi R^{2}} \label{FofR},
\end{equation}
where $R$ is the distance between the quasar and the point of interest
%$d$ is the luminosity distance to the quasar, $z$ is the quasar redshift,
 and $L_{\mathrm{\nu_{L}}}$ is the luminosity at the Lyman limit.
Combining equation (\ref{gammaf}) and (\ref{FofR}), we have %and using the UV background photoionization rate $\Gamma_{\mathrm{bkg}}$,
%the function $\Gamma_{\mathrm{Q}}$ of $R$ can be written as %from the definition of the proximity zone as
\begin{equation}
\Gamma_{\mathrm{Q}}(R) = \frac{9.5 \times10^{8}}{\alpha+2.75}\frac{L_{\mathrm{\nu_{L}}}}{4 \pi R^{2}}. \label{calc} %= \Gamma_{\mathrm{bkg}}.
\end{equation}
Then the proximity zone size $R_{\mathrm{pz}}$ is defined as the distance from the quasar to the point where the ionization rate of the quasar radiation is equal to that of the background radiation, $\Gamma_{bkg}$:
\begin{equation}
\Gamma_{\mathrm{Q}}(R_{\mathrm{pz}}) = \Gamma_{\mathrm{bkg}} \label{def}
\end{equation}
% If we change equation (\ref{A}) to the form of $R_{\mathrm{pz}}$
From equation (\ref{calc}) and (\ref{def}), we obtain %the proximity size $R_{\mathrm{pz}}$ as follows.
\begin{equation}
R_{\mathrm{pz}} = \biggl(\frac{9.5 \times10^{8}}{4\pi(\alpha+2.75)}\frac{L_{\mathrm{\nu_{L}}}}{\Gamma_{\mathrm{bkg}}}\biggr)^\frac{1}{2}.
\end{equation}
 In the estimation of the proximity zone size, the quasar continuum is fitted in the wavelength windows of $960-1100$ \AA \, and $1700-1900$ \AA \, to obtain $\alpha$.
 We measure the average of the luminosity at $912-940$\,\AA \, as $L_{\mathrm{\nu L}}$ and use the value of cosmic background photoionization rate $\Gamma_{\mathrm{bkg}} = 0.792 \times 10^{-12} \mathrm{s^{-1}}$ at $z\sim3$ \citep{Haardt2012}.
 %To check the validity of the estimation of proximity zone size, 
%we draw the relation between the BH masses and the proximity zone sizes (Figure \ref{prop}). 
% to check the certainity of proximity zone size in Figure \ref{prop}.
Figure \ref{prop} presents a clear correlation between the measured black hole masses and proximity zone sizes, both of which trace the accumlated past quasar activity.
The correlation is fairly strong, with the Spearman's rank correlation coefficient and $p$-value of $0.69$ and $5.8\times10^{-10}$ ($\sim 6.1 \sigma$), respectively.
It is unlikely that selection bias due to selection effects that tend to miss more luminous quasars at a fixed black hole mass.

%%%%%%%%%%%%%%%%%%%%%%%%%%%%%%%%%%%%%%%%%%%%%%%%%%%%%%%%%%%%%%%%%%%%%%%%%%%%%%%%%%%%%%%%%%%%%%%%%%%%%%%
%%%%%%%%%%%%%%%%%%%%%%%%%%%%%%%%%%%%%%%%%%%%%%%%%%%%%%%%%%%%%%%%%%%%%%%%%%%%%%%%%%%%%%%%%%%%%%%%%%%%%%%

%%%%%%%%%%%%%%%%%%%%%%%%%%%%%%%%%%%%%%%%%%%%%%%%%%%%%%%%%%%%%%%%%%%%%%%%%%%%%%%%%%%%%%%%%%%%%%%%%%%%%%%
\begin{figure*}[t]
\begin{center}
\includegraphics[width=1.1\linewidth]{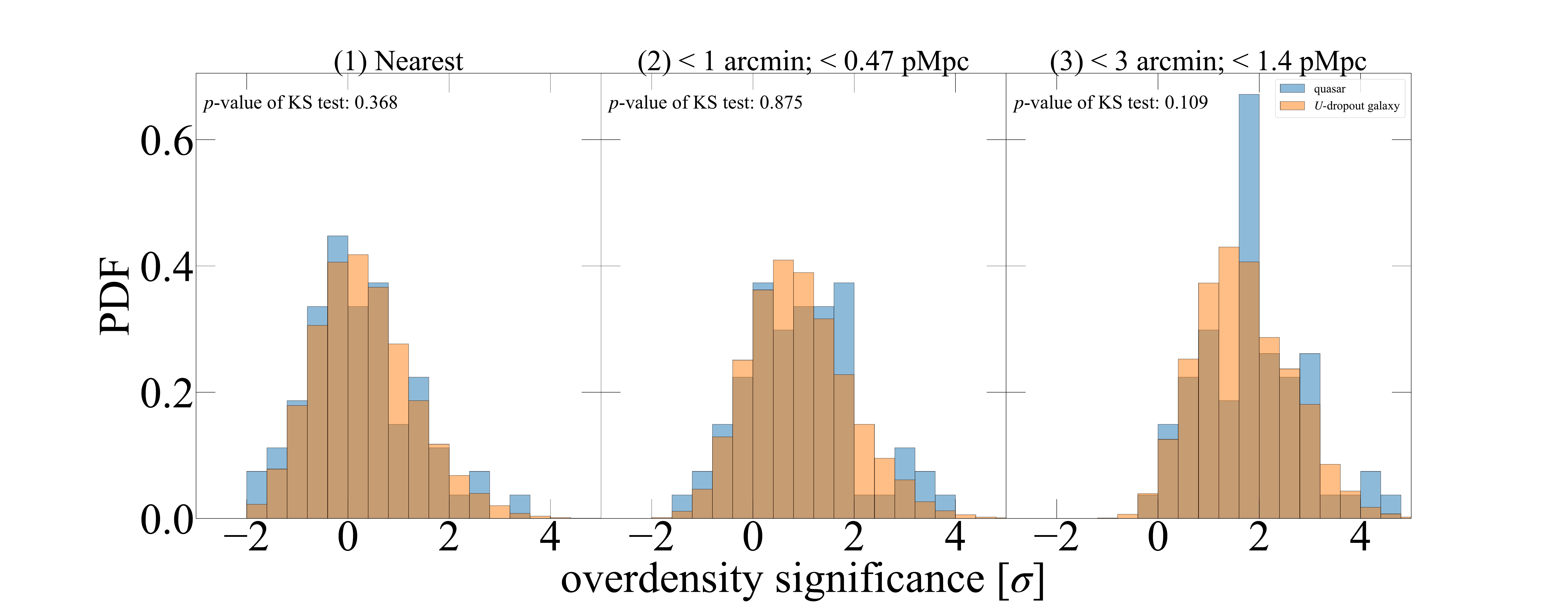}
\end{center}
\vspace*{-0.7cm}
\caption{
The probability density function (PDF) of overdensity significances around the quasars (blue bars) and around the control sample of $U$-dropout galaxies (orange bars). 
Panel (1), (2), and (3) present the cases of Nearest, $<1$ arcmin, and $<3$ arcmin, respectively. 
%Panel (1) shows a histogram of the overdensity significance located on the quasars and Panel (2) and (3) show histograms of the maximum values of overdensity significances on the quasars with radius 1 arcmin and 3 arcmin. 
}  \label{histgram}
\end{figure*}
%%%%%%%%%%%%%%%%%%%%%%%%%%%%%%%%%%%%%%%%%%%%%%%%%%%%%%%%%%%%%%%%%%%%%%%%%%%%%%%%%%%%%%%%%%%%%%%%%%%%%%%

%%%Result
\section{Results and Discussion} \label{sec:results}
%\subsection{The overdensities around quasars} \label{subsec:tables}
Figure \ref{histgram} shows the probability distributions of the overdensity significances around the quasars and the control sample of $U$-dropout galaxies.
The distributions are similar to each other at all the three scales.
The $p$-values of the Kolmogorov-Smirnov (KS) test between the two distributions are 0.368, 0.875, and 0.109 for the Nearest, $<1$ arcmin, and $<3$ arcmin cases, respectively, significantly larger than 0.05 below which the two samples are considered to be drawn from different distributions.
% This shows that the two distributions cannot be said as different.
%We also find that 3 out of the 67 quasars are associated with protocluster candidates within projected separation of 3 arcmin.
%This is further discussed in Section \ref{subsec:dmhalo}. 

%%%%%%%%%%%%%%%%%%%%%%%%%%%%%%%%%%%%%%%%%%%%%%%%%%%%%%%%%%%%%%%%%%%%%%%%%%%%%%%%%%%%%%%%%%%%%%%%%%%%%%%
\begin{figure*}[t]
\begin{center}
\includegraphics[width=1.1\linewidth]{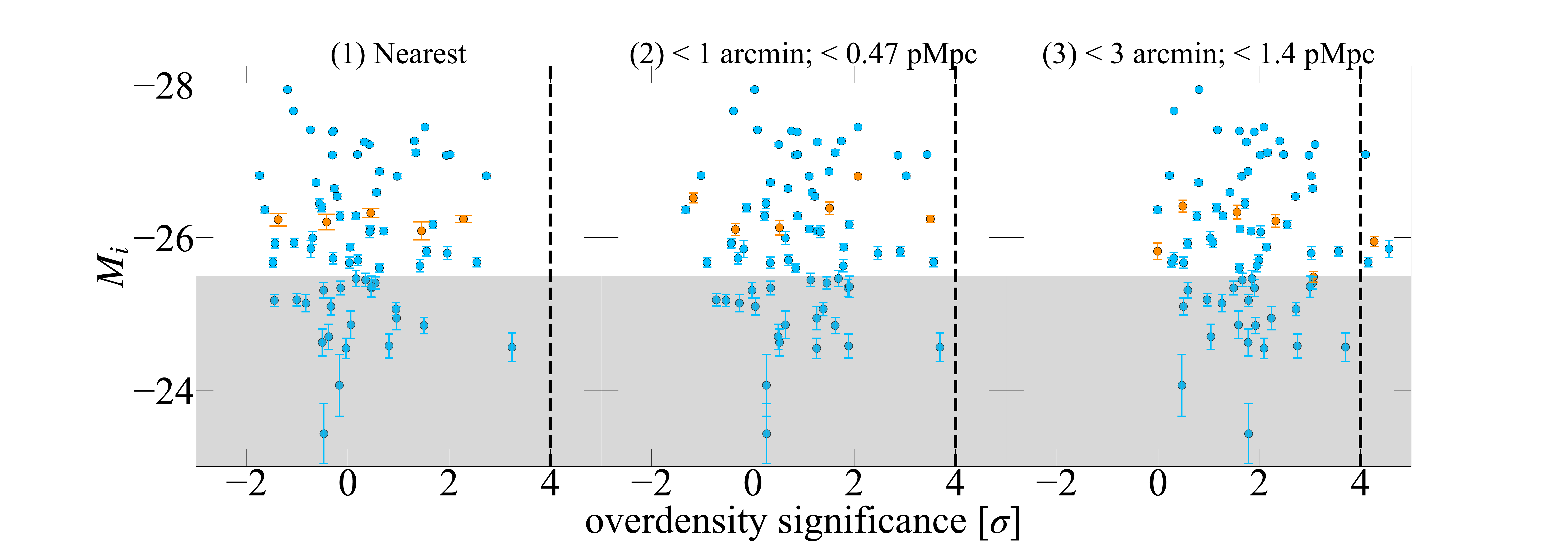}
\end{center}
\vspace*{-0.7cm}
\caption{
The quasar UV absolute magnitudes, $M_i$, versus the overdensity significances around the quasars. 
Panel (1), (2), and (3) show the cases of Nearest, $<1$ arcmin, and $<3$ arcmin, respectively. 
The median values of the complete sample in bins of the overdensity significance are indicated by the orange dots, with error bars representing the standard error of the median.
The dashed vertical lines indicate the $4\sigma$ overdensity, above which the region is considered to be a protocluster. 
%The gray shaded region shows the complete limit.
The quasar sample is not complete in the gray shaded parameter space.}
\label{UV}
% that corresponds to the eBOSS limiting magnitude of $r<22$ and assuming that $r-i=0.0954$ which is the median color of the our quasar samples.
\end{figure*}

%%%%%%%%%%%%%%%%%%%%%%%%%%%%%%%%%%%%%%%%%%%%%%%%%%%%%%%%%%%%%%%%%%%%%%%%%%%%%%%%%%%%%%%%%%%%%%%%%%%%%%%

%%%%%%%%%%%%%%%%%%%%%%%%%%%%%%%%%%%%%%%%%%%%%%%%%%%%%%%%%%%%%%%%%%%%%%%%%%%%%%%%%%%%%%%%%%%%%%%%%%%%%%%

\begin{figure*}[t]
\begin{center}
\includegraphics[width=1.1\linewidth]{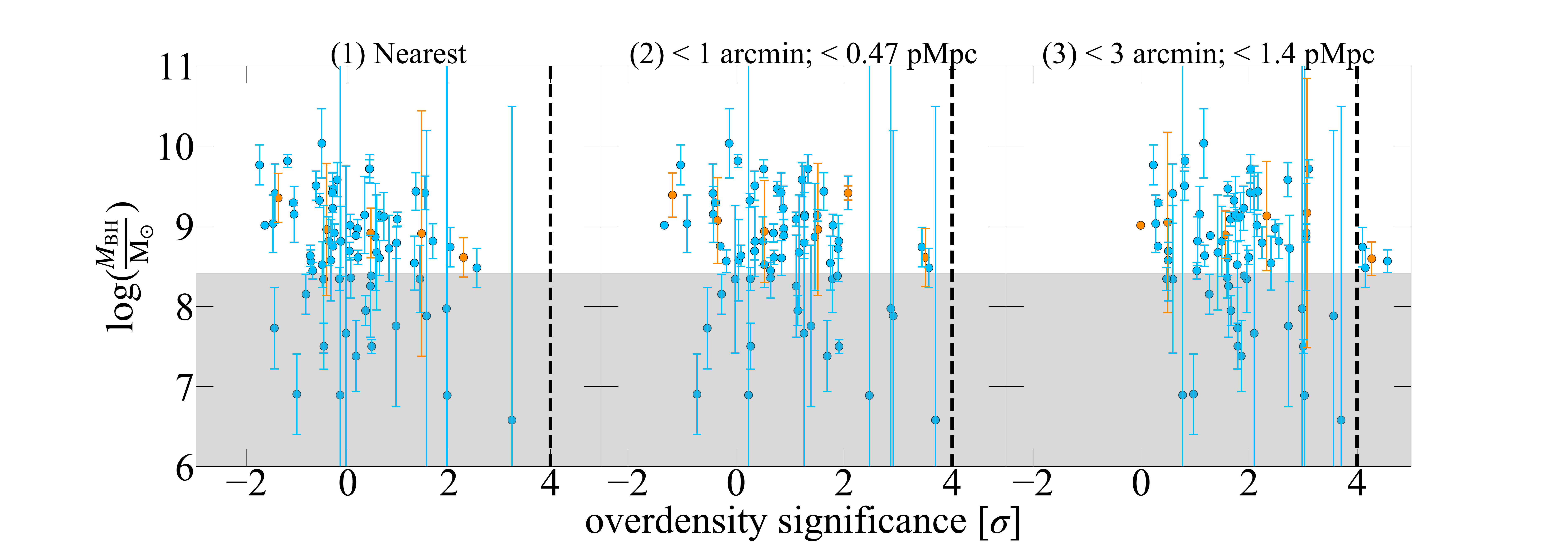}
\end{center}
\vspace*{-0.7cm}
\caption{
%Identical to Figure \ref{UV}, but for
The quasar BH masses versus the overdensity significance.
The symbols are the same as in Figure 3.
} 
\label{BH}
\end{figure*}

%%%%%%%%%%%%%%%%%%%%%%%%%%%%%%%%%%%%%%%%%%%%%%%%%%%%%%%%%%%%%%%%%%%%%%%%%%%%%%%%%%%%%%%%%%%%%%%%%%%%%%%

%%%%%%%%%%%%%%%%%%%%%%%%%%%%%%%%%%%%%%%%%%%%%%%%%%%%%%%%%%%%%%%%%%%%%%%%%%%%%%%%%%%%%%%%%%%%%%%%%%%%%%%
\begin{figure*}[t]
\begin{center}
\includegraphics[width=1.1\linewidth]{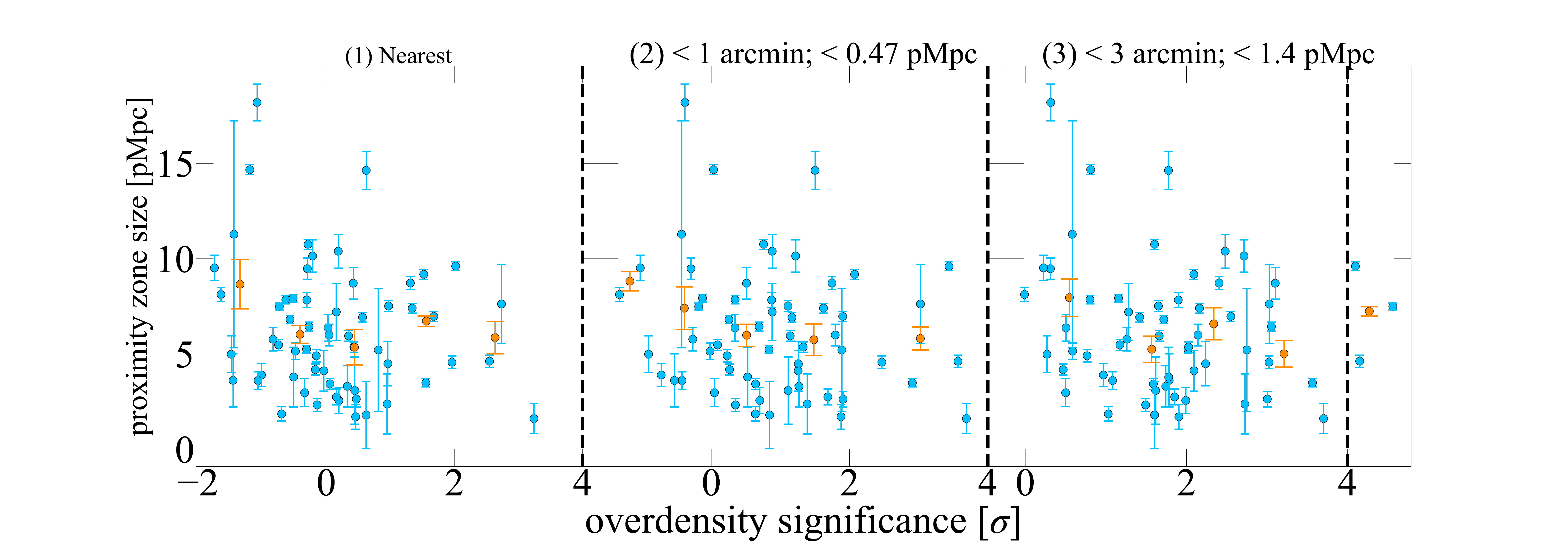}
\end{center}
\vspace*{-0.7cm}
\caption{
The proximity zone sizes versus the overdensity significance.
The symbols are the same as in Figure 3.
}
 \label{pz}
\end{figure*}
%%%%%%%%%%%%%%%%%%%%%%%%%%%%%%%%%%%%%%%%%%%%%%%%%%%%%%%%%%%%%%%%%%%%%%%%%%%%%%%%%%%%%%%%%%%%%%%%%%%%%%%

Figures \ref{UV} - \ref{pz} present correlations between quasar properties (UV luminosity, BH mass, and proximity zone size) and the overdensity significance, and Table \ref{table1} lists the results of Spearman's correlation tests for the all and complete samples.
None of the listed properties are significantly correlated with the overdensity significance at all scales.
We will discuss these results in Section \ref{subsec:correlation}.

%%%%%%%%%%%%%%%%%%%%%%%%%%%%%%%%%%%%%%%%%%%%%%%%%%%%%%%%%%%%%%%%%%%%%%%%%%%
\begin{table}[h!]
 \caption{Results of Spearman's rank correlation tests for overdensity signicance and quasar properties}
 \centering
 \begin{tabular*}{8.0cm}{@{\extracolsep{\fill}}ccc}
 \tableline
 \tableline
 UV luminosity (all sample) & $\rho^{\ast}$ & $p$-value\\
 \tableline
 \tableline
 Nearest & 0.010 & 0.93 \\
 \textless~1 arcmin & -0.023 & 0.85 \\
 \textless~3 arcmin & -0.013 & 0.91 \\
 \tableline
 UV luminosity (complete sample) & $\rho^{\ast}$ & $p$-value\\
 \tableline
 Nearest & 0.067 & 0.66 \\
 \textless~1 arcmin & -0.030 & 0.84 \\
 \textless~3 arcmin & -0.052 & 0.73 \\
 \tableline
 black hole mass (all sample) & $\rho^{\ast}$ & $p$-value\\
 \tableline
 Nearest & -0.24 & 0.052 \\
 \textless~1 arcmin & -0.20 & 0.10 \\
 \textless~3 arcmin & -0.19 & 0.13 \\
 \tableline
 black hole mass (complete sample) & $\rho^{\ast}$ & $p$-value\\
 \tableline
 Nearest & -0.23 & 0.12 \\
 \textless~1 arcmin & -0.16 & 0.29 \\
 \textless~3 arcmin & -0.15 & 0.34 \\
 \tableline
 proximity zone size& $\rho^{\ast}$ & $p$-value\\
 \tableline
 Nearest & -0.17 & 0.20 \\
 \textless~1 arcmin & -0.17 & 0.20 \\
 \textless~3 arcmin & -0.10 & 0.44 \\
 \tableline
 \tableline
\end{tabular*}
\label{table1}
\end{table}
%%%%%%%%%%%%%%%%%%%%%%%%%%%%%%%%%%%%%%%%%%%%%%%%%%%%%%%%%%%%%%%%%%%%%%%%%%%

%%% Discussion
%\section{Discussion} \label{sec:discussion}

\subsection{Quasar DM halo mass} \label{subsec:dmhalo}
Three out of the 67 quasars were found to associate with $>4\sigma$ overdense regions.
%In section 4.1, three out of 67 quasars associate with the $>4\sigma$ overdense regions.
%To confirm on the aspect of the quasar halo mass, we check how the halo masses at $z\sim3$ evolve to $z=0$.
%%
%We confirm that this ratio is consistent with an aspect of quasar halo masses by comparing it with the halo mass estimated by \citet{Shen2007} and \citet{Eftekharzadeh2015} \citep{Uchiyama2018, Uchiyama2020} as follow.
Here we estimate the mass of the quasar dark matter haloes from this fraction using a theoretical model, and compare it with those estimated by \citet{Shen2007} and \citet{Eftekharzadeh2015}. %\citep{Uchiyama2020}.
When we take into account the purity $\sim76 \pm 15\%$ and the completeness $\sim5 \pm 1 \%$ of the protocluster candidates (Section \ref{subsec:overdensity}), $(3^{+2.9}_{-1.4})\times(0.76\pm0.15)/(0.05\pm0.01) = 46^{+44}_{-32}$ quasars are expected to be associated with protoclusters \citep[the error comes from Poisson statistics;][]{Gehrels1986}. %will be $>10^{14}\,\mathrm{M_{\odot}}$ at $z=0$.
%With the estimated , it indicates that $(3\pm1.7)\times0.81 $ quasars are actually associated with protoclusters.
%Also, with the completeness $\sim5\%$ of the protocluster candidate \citep{Toshikawa2016}, 48.6 of the quasars associated with the region that are $>10^{14}\,\mathrm{M_{\odot}}$ at $z=0$.
Thus, $69^{+31}_{-48}\%$ of the quasars in this study are likely hosted by haloes that will evolve to $>10^{14}\,\mathrm{M_{\odot}}$ at $z=0$.
With this fraction, we estimate the halo mass at $z=3.1$ by using the extended Press-Schechter (EPS) model \citep{Bond1991, Bower1991, Lacey1993}.
%found in \citet{Shen2007} and \citet{Eftekharzadeh2015}.
In the EPS model, we can estimate a conditional probability $P_{2}(M_{\mathrm{t2}},z_{2}|M_{\mathrm{t1}},z_{1})$ that a halo mass $M_{\mathrm{t1}}$ at a redshift $z_{1}$ evolves to a halo mass $M_{\mathrm{t2}}$ at a later redshift $z_{2}$ \citep{Hamana2006}.
Subsequently, the conditional mass function $n_{2}(M_{\mathrm{t2}},z_{2}|M_{\mathrm{t1}},z_{1})$ is estimated as follows \citep{Hamana2006}:
\begin{equation}
n_{2}(M_{\mathrm{t2}},z_{2}|M_{\mathrm{t1}},z_{1})dM_{\mathrm{t2}} \propto \frac{1}{M_{\mathrm{t2}}} P_{2}(M_{\mathrm{t2}},z_{2}|M_{\mathrm{t1}},z_{1})dM_{\mathrm{t2}}.
\end{equation}
Here, we search for possible values of $M_{\mathrm{t1}}$ at $z_{1}=3.1$ that has the $69^{+31}_{-48}\%$ probability of having $M_{\mathrm{t2}}>10^{14}\,\mathrm{M_{\odot}}$ at $z_{2}=0$,
 %On the other hand,  estimated the average halo mass of $0.66^{+0.06}_{-0.06} \times 10^{12}h^{-1}M_{\odot}$ at $z\sim3.0$.
and find that $M_{\mathrm{t1}}=1.3^{+1.4}_{-0.9} \times 10^{13} h^{-1} \mathrm{M_{\odot}}$ meets this condition. %at $z_{1}=3.1$ will be $>10^{14}\,\mathrm{M_{\odot}}$ at $z_{2}=0$.
On the other hand, \citet{Shen2007} and \citet{Eftekharzadeh2015} estimated the average halo mass of about $5.0^{+3.0}_{-2.0} \times 10^{12}h^{-1}\mathrm{M_{\odot}}$ at $z=3.2$ and $0.66^{+0.06}_{-0.06} \times 10^{12}h^{-1}\mathrm{M_{\odot}}$ at $z=3.0$, respectively.
Our results are therefore closer to that of \citet{Shen2007} but are also broadly consistent with \citet{Eftekharzadeh2015}.

\subsection{Correlation between quasar properties and local densities} \label{subsec:correlation}
In Figures \ref{UV} - \ref{pz}, we found no correlation between the quasar properties and the surrounding density environments. 
On the other hand, \citet{Uchiyama2018} found that the UV luminosities, the BH masses, and the near zone sizes of quasars tend to correlate with the overdensity significances only at the points where quasars reside in (i.e., ``Nearest'' in this study) at $z\sim4$, suggesting the photoevaporation or other forms of quasar feedback at work, but not on larger scales.
% This suggests feedback from quasars to surrounding galaxies, such as photoevaporation.
Since the sample size of this study is smaller than that of \citet{Uchiyama2018}, it is possible that correlations are not seen even at the smallest scale.

\begin{figure*}[t]
\begin{center}
\includegraphics[width=1.1\linewidth]{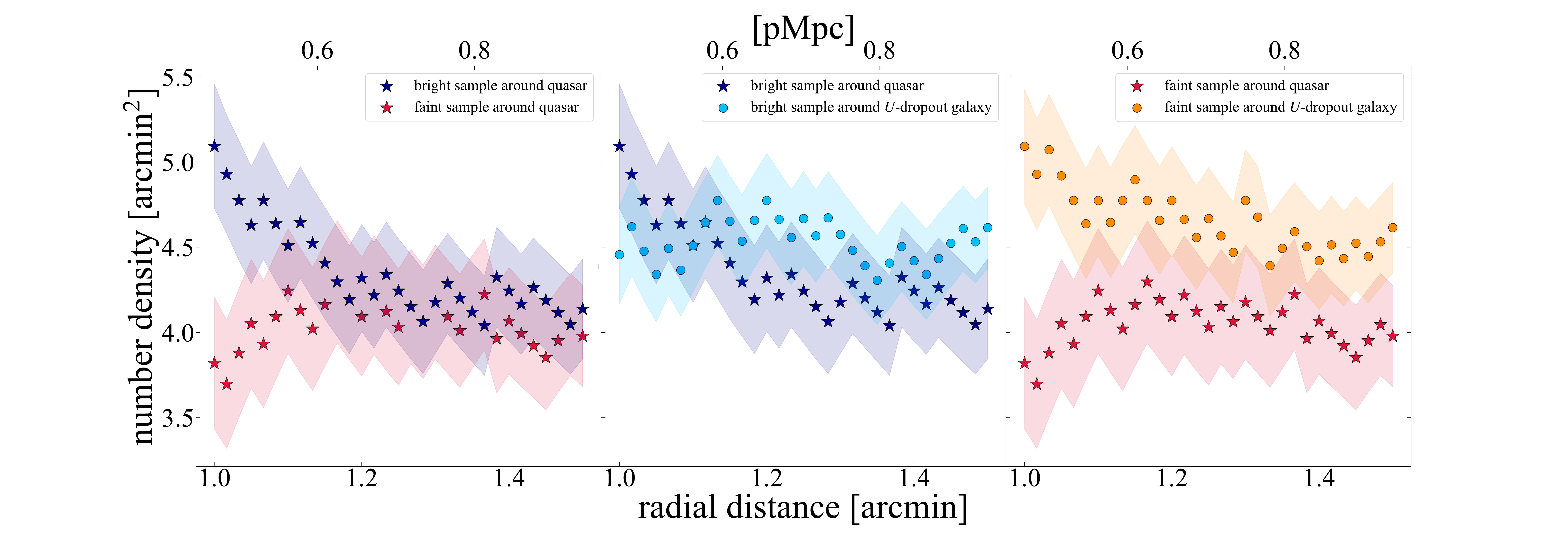}
\end{center}
\vspace*{-0.8cm}
\caption{Radial profile of the median stacked number density of galaxies.
%The stars (points) show the case of the density around quasars ($u$-dropout galaxies).
The stars and dots represent the density of bright (blue/lightblue) and faint (red/orange) galaxies around the quasars/control sample of galaxies, respectively.
%The blues (oranges) represent the density of luminous (faint) ones.
The shaded regions indicate the 1$\sigma$ error.
Left: comparison of the density profiles of bright and faint $U$-dropout galaxies around quasars.
Middle: comparison of the density profiles of bright $U$-dropout galaxies around quasars and around $U$-dropout galaxies.
Right: comparison of the density profiles of faint $U$-dropout galaxies around quasars and around $U$-dropout galaxies.}
\label{radial}
\end{figure*}

In order to resolve such proximity effects, we plot in Figure \ref{radial} the radial density of $U$-dropout galaxies profiles within the projected distances of $<1.6$ arcmin centered on the quasars. %(see also Figure \ref{picture}, which shows a image around our quasar by hscmap to confirm the scale).}
Here, we divide the surrounding galaxy sample into the ``bright'' and ``faint'' subsamples at the median magnitude, $r=25.9$.  %which is the median magnitude. 
%The subsamples with $r<25.9$ and $r\ge25.9$ are called to ``luminous galaxies" and ``faint galaxies", respectively. 
The median magnitudes of the bright and faint galaxies are $r=25.4$ and $r=26.2$, respectively. 
In the left panel of Figure \ref{radial}, we find that the density of fainter galaxies decreases as they get closer to the quasar at $<1.2$ arcmin.
We fit two straight lines to the data points of the faint galaxies in this panel, one at $<1.2$ arcmin and the other at $>1.2$ arcmin.
From the intersection of the two lines, we determine that $0.51\pm0.05$ pMpc ($= 1.1\pm0.1$ arcmin) is the distance at which the density of the faint galaxies begins to decrease toward the quasars, while the density is almost flat at the larger distances.
%As shown in the figure, we find that the median density of the faint galaxies is lower than that of the bright galaxies within the projected distances of 0.46 pMpc from the quasars by about 2 $\sigma$ level. 
The middle panel of Figure \ref{radial} shows that the densities of the bright galaxies around the quasars are comparable to those of the bright galaxies around the $U$-dropout galaxies.
In the right panel, we find that the density of the faint galaxies is lower around the quasars than around the galaxies.
Especially, in the vicinity of the quasars, a similar trend can be seen at the same scale as in the left panel.

%%%%%%%%%%%%%%%%%%%%%%%%%%%%%%%%%%%%%%%%%%%%%%%%%%%%%%%%%%%%%%%%%%%%%%%%%%%%%%%%%%%%%%%%%%%%%%%%%%%%%%%
%%%%%%%%%%%%%%%%%%%%%%%%%%%%%%%%%%%%%%%%%%%%%%%%%%%%%%%%%%%%%%%%%%%%%%%%%%%%%%%%%%%%%%%%%%%%%%%%%%%%%%%

%the density of the faint galaxies is lower only in the vicinity of the quasar (left bottom panel). 

We check whether this segregation of the surrounding density is caused by the photoevaporation effect by the quasar radiation. 
It is evaluated by comparing the UV intensity of the central quasar to the UV background radiation at the Lyman limit \citep[e.g.,][]{Kashikawa2007, Uchiyama2019}.
The isotropic UV intensity $J_{\mathrm{\nu}}$ from the central quasar is expressed by the following formula \citep{Kashikawa2007, Uchiyama2019, Santos2022}:
\begin{equation}
J_{\mathrm{\nu}} = J_{\mathrm{21}}\biggl(\frac{\nu}{\nu_{\mathrm{L}}}\biggr)^{\alpha} \times 10^{-21} \, \mathrm{erg\,cm^{-2}\,s^{-1}\,Hz^{-1}\,sr^{-1}},
\end{equation}
where $J_{21}$ is the intensity at the Lyman limit. %, $\nu_{\mathrm{L}}$ is the frequency at the Lyman limit, and $\alpha$ is the slope of quasar continuum.
Since $J_{\mathrm{\nu}}=L_{\mathrm{\nu}}/(4 \pi r)^{2}$, we have 
\begin{equation}
r=\frac{1}{4\pi} \sqrt{\frac{L_{\mathrm{\nu_{L}}}}{J_{\mathrm{21}}(\frac{\nu}{\nu_{\mathrm{L}}})^{\alpha} \times 10^{-21}}}.
\end{equation}
%where $L_{\mathrm{\nu_{L}}}$ is the luminosity at the Lyman limit.
When the intensity of quasar UV radiation $J_{\mathrm{21}}$ is equal to that of UV background radiation, $r$ is equivalent to the proximity zone size.
The distance $r=0.51\pm0.05$ pMpc corresponds to the quasar UV intensity at the Lyman limit of $J_{21}=27\pm5.4$.
On the other hand, the intensity at the Lyman limit of the UV background radiation at $z=3$ is evaluated to be $J_{\mathrm{21}}=1.0^{+0.5}_{-0.3}$ from the quasar proximity effect measurements \citep{Cooke1997}.
Thus, it seems that the density of the faint galaxies is reduced around the quasars where the quasar UV radiation is $\sim30$ times stronger than the background radiation.
%We find that quasar radiation is more than 30 times stronger than the UV background radiation within $r \sim 0.46 \, \mathrm{pMpc}$.
%We find the radius where $J_{21}=30$, i.e., $r_{\mathrm{prox}} \sim 0.46 \, \mathrm{pMpc}$ for  where the local UV radiation is 30 times greater than UV background radiation.
%This corresponds to the scale where the density of faint galaxies is significantly smaller in left panel of Figure \ref{radial}.
%This suggests that the photoevaporation may act on the faint $U$-dropout galaxies.
%

\citet{Harikane2022} estimated the halo mass of $U$-dropout galaxies, in essentially the same luminosity range as our $U$-dropout galaxies, to be $\sim 2.2 \times 10^{11} \, \mathrm{M_{\odot}}$. %at the less massive end. 
\citet{Kashikawa2007} suggests that quasar radiation at Lyman limit of $J_{21}=30$ can suppress star fomation in a halo with a virial mass of $M_{\mathrm{vir}}< 10^{10} \, \mathrm{M_{\odot}}$, but the halo mass of our sample is significantly larger.
%The quasar UV enhanced region reaches $\sim 0.46 \, \mathrm{pMpc}$ when quasar UV radiation is 30 times greater than UV background radiation. 
%Thus, our result of the density segregation is suggested to be caused by not the photoevaporation effects but the other baryionic physics. 
\citet{Santos2022} found a similar result in the case of LAE environments.
The brighter LAEs prefer to reside in more massive DM haloes and should be expected to be in denser regions but they reported the opposite trend, with fainter LAEs avoiding the bright central LAEs.
This may imply that UV radiation from the central LAEs inhibits the formation of surrounding LAEs.
They estimated the $J_{21}$ of the central LAEs and compared the mass of the surrounding halo with the virial mass at which star formation is suppressed, based on \citet{Kashikawa2007}.
While they found that the UV radiation of central LAEs can suppress the star formation at $M_{\mathrm{vir}}<10^{9}\, \mathrm{M_{\odot}}$, the average halo mass of the LAEs is $M_{\mathrm{vir}}\sim10^{10-11} \mathrm{M_{\odot}}$ measured by clustering analysis in \citet{Ouchi2018}, indicating that star formation can be suppressed at such larger haloes.
Galaxies in a massive halo may thus be affected by photoevaporation, depending on the central UV radiation and property of the surrounding galaxies.
%Based on the discussion so far, it may be that smaller scales, such as $<$1 pMpc, are more relevant and important than protocluster scale ($>1$ pMpc) for the environment around quasars.
%However, the redshift range of our $U$-dropout galaxy sample is $\Delta z \sim 0.5$ \citep{Toshikawa2024}.
%It should be noted that the densities are the results of the contaminating the foreground/background galaxies.

\section{Conclusion}\label{subsec:tables}
We have measured the local density environments around SDSS luminous quasars at $z\sim3$ using the imaging data from HSC-SSP and CLAUDS. 
We measured the $U$-dropout galaxy densities around 67 quasars, and found that the density distributions of the quasars and the $U$-dropout galaxies are comparable.
Three quasars are associated with protocluster candidates within a separation of 3 arcmin.
The fraction of such association suggests that quasars at this epoch occupy haloes with a mass of $1.3^{+1.4}_{-0.9}\times 10^{13} h^{-1} \mathrm{M_{\odot}}$, which is consistent with a measurement by \citet{Shen2007}.
We also analyzed the correlation between the surrouding densities and quasar properties (UV luminosity, BH mass, and proximity zone size). % as found in \citet{Uchiyama2018}.
%The analysis showed that
As a result, no significant correlations are found.
%\citet{Uchiyama2018} found the correlations only at the points where quasars reside in, thus
Finally, we examined the radial density profile of galaxies around the quasars.
% on a smaller scale to resolve the proximity effects.
The local density of faint $U$-dropout galaxies are found to be significantly lower than that of bright $U$-dropout galaxies within a projected distance of $0.51\pm0.05$ pMpc from the quasars.
This distance corresponds the quasar UV intensity at the Lyman limit of $J_{21} = 27 \pm 5.4$, which is about 30 times that of the background radiation.
%the local density 
Photoevaporation may suppress galaxy formation at short distances where the quasar UV intensity is strong, even in massive haloes.

\vskip\baselineskip
%\begin{acknowledgments}
\section*{Acknowledgments}
Y.S. was supported by JST, the establishment of university fellowships towards the creation of science technology innovation, Grant Number JPMJFS2131 and JST SPRING, Japan Grant Number JPMJSP2162.
Y.M. was supported by the Japan Society for the Promotion of Science (JSPS) KAKENHI grant No. 21H04494.
C.L. acknowledges support from the National Natural Science Foundation of China (NSFC,
Grant No. 11933003), 111 project (No. B20019), and Key Laboratory for Particle
Physics, Astrophysics and Cosmology, Ministry of Education.

The Hyper Suprime-Cam (HSC) collaboration includes the astronomical communities of Japan and Taiwan, and Princeton University.
The HSC instrumentation and software were developed by the National Astronomical Observatory of Japan (NAOJ), the Kavli Institute for the Physics and Mathematics of the Universe (Kavli IPMU), the University of Tokyo, the High Energy Accelerator Research Organization (KEK), the Academia Sinica Institute for Astronomy and Astrophysics in Taiwan (ASIAA), and Princeton University.
Funding was contributed by the FIRST program from Japanese Cabinet Office, the Ministry of Education, Culture, Sports, Science and Technology (MEXT), the Japan Society for the Promotion of Science (JSPS), Japan Science and Technology Agency (JST), the Toray Science Foundation, NAOJ, Kavli IPMU, KEK, ASIAA, and Princeton University. 
 This paper makes use of software developed for the Large Synoptic Survey Telescope.
We thank the LSST Project for making their code available as free software at  http://dm.lsst.org 
The Pan-STARRS1 Surveys (PS1) have been made possible through contributions of the Institute for Astronomy, the University of Hawaii, the Pan-STARRS Project Office, the Max-Planck Society and its participating institutes, the Max Planck Institute for Astronomy, Heidelberg and the Max Planck Institute for Extraterrestrial Physics, Garching, The Johns Hopkins University, Durham University, the University of Edinburgh, Queen's University Belfast, the Harvard-Smithsonian Center for Astrophysics, the Las Cumbres Observatory Global Telescope Network Incorporated, the National Central University of Taiwan, the Space Telescope Science Institute, the National Aeronautics and Space Administration under Grant No. NNX08AR22G issued through the Planetary Science Division of the NASA Science Mission Directorate, the National Science Foundation under Grant No. AST-1238877, the University of Maryland, and Eotvos Lorand University (ELTE) and the Los Alamos National Laboratory. 
Based on data collected at the Subaru Telescope and retrieved from the HSC data archive system, which is operated by Subaru Telescope and Astronomy Data Center at National Astronomical Observatory of Japan.
This work is based on data collected at the Subaru Telescope and retrieved from the HSC data archive system, which is operated by the Subaru Telescope and Astronomy Data Center at the National Astronomical Observatory of Japan.

These data were obtained and processed as part of the CFHT Large Area U-band Deep Survey (CLAUDS), which is a collaboration between astronomers from Canada, France, and China described in \citet{Sawicki2019}. %, [MNRAS 489, 5202]).
CLAUDS is based on observations obtained with MegaPrime/ MegaCam, a joint project of CFHT and CEA/DAPNIA, at the CFHT which is operated by the National Research Council (NRC) of Canada, the Institut National des Science de l'Univers of the Centre National de la Recherche Scientifique (CNRS) of France, and the University of Hawaii.
CLAUDS uses data obtained in part through the Telescope Access Program (TAP), which has been funded by the National Astronomical Observatories, Chinese Academy of Sciences, and the Special Fund for Astronomy from the Ministry of Finance of China.
CLAUDS uses data products from TERAPIX and the Canadian Astronomy Data Centre (CADC) and was carried out using resources from Compute Canada and Canadian Advanced Network For Astrophysical Research (CANFAR).
%\end{acknowledgments}

%\section*{References}
\bibliographystyle{aasjournal}
\bibliography{reference}

\end{document}